\documentclass[conference]{IEEEtran}
\IEEEoverridecommandlockouts

\usepackage[T1]{fontenc}%
\usepackage{graphicx}
\usepackage{bm}
\usepackage{hyperref}
\usepackage{caption}
\usepackage{subcaption}
\usepackage{booktabs}
\usepackage{multirow}
\usepackage{siunitx}
\usepackage{pifont}%
\newcommand{\cmark}{\ding{51}}%
\newcommand{\xmark}{\ding{55}}%

\usepackage{amsmath,amsfonts,bm}

\def\eqref#1{equation~\ref{#1}}

\def\1{\bm{1}}

\def\rvx{{\mathbf{x}}}
\def\rvy{{\mathbf{y}}}

\def\rmH{{\mathbf{H}}}

\def\vh{{\bm{h}}}

\def\vx{{\bm{x}}}
\def\vy{{\bm{y}}}
\def\vz{{\bm{z}}}

\def\mI{{\bm{I}}}

\DeclareMathAlphabet{\mathsfit}{\encodingdefault}{\sfdefault}{m}{sl}
\SetMathAlphabet{\mathsfit}{bold}{\encodingdefault}{\sfdefault}{bx}{n}

\def\gN{{\mathcal{N}}}

\def\gX{{\mathcal{X}}}

\def\sC{{\mathbb{C}}}

\newcommand{\E}{\mathbb{E}}

\newcommand{\R}{\mathbb{R}}
\newcommand{\C}{\mathbb{C}}

\usepackage{todonotes}

\usepackage{color, soul}
\usepackage[printonlyused]{acronym}
\acrodef{AR}{auto-regressive}
\acrodef{ARMA}{auto-regressive moving average}
\acrodef{CP}{cyclic prefix}
\acrodef{CDL}{clustered delay line}
\acrodef{CIR}{channel impulse response}
\acrodef{KF}{Kalman Filter}
\acrodef{MMSE}{Minimum Mean Square Error}
\acrodef{MNSE}{Mean Normalized Square Error}
\acrodef{GAN}{generative adversarial network}

\acrodef{NSE}{Normalized Square Error}
\acrodef{NN}{Neural Network}
\acrodef{HKF}{Hypernetwork Kalman Filter}
\acrodef{BKF}{Binned Kalman Filter}
\acrodef{GKF}{genie \ac{KF}}
\acrodef{UE}{User Equipment}
\acrodef{ISI}{Intersymbol Interference}

\makeatletter
\DeclareRobustCommand\bfseries{%
  \not@math@alphabet\bfseries\mathbf
  \fontseries\bfdefault\selectfont
  \boldmath %
}
\makeatother

\newcommand{\myparagraph}[1]{
\vspace{0.20cm}\noindent
\textbf{#1.}
}

\begin{document}
\title{MIMO-GAN: Generative MIMO Channel Modeling}
\author{
    \IEEEauthorblockN{Tribhuvanesh Orekondy\IEEEauthorrefmark{1}, 
    Arash Behboodi\IEEEauthorrefmark{1}, 
    Joseph B. Soriaga\IEEEauthorrefmark{2} 
    }
    \IEEEauthorblockA{\IEEEauthorrefmark{1}Qualcomm Technologies Netherlands B.V.}
    \IEEEauthorblockA{\IEEEauthorrefmark{2}Qualcomm Technologies, Inc.}
    \IEEEauthorblockA{Qualcomm AI Research} 
\thanks{Qualcomm AI Research is an initiative of Qualcomm Technologies, Inc.}
}

\maketitle

\begin{abstract}
We propose generative channel modeling to learn statistical channel models from channel input-output measurements. Generative channel models can learn more complicated distributions and represent the field data more faithfully. They are tractable and easy to sample from, which can potentially speed up the simulation rounds. To achieve this, we leverage advances in \ac{GAN}, which helps us learn an implicit distribution over stochastic MIMO channels from observed measurements. 
In particular, our approach MIMO-GAN implicitly models the wireless channel as a distribution of time-domain band-limited impulse responses. 
We evaluate MIMO-GAN on 3GPP TDL MIMO channels and observe high-consistency in capturing power, delay and spatial correlation statistics of the underlying channel.
In particular, we observe MIMO-GAN achieve errors of under 3.57 ns average delay and -18.7 dB power.
\end{abstract}

\section{Introduction}
\label{sec:intro}

Statistical channel models, from Rayleigh and Rician fading models to 3GPP models like TDL and CDL \cite{3gpp_38901_2020}, are widely used for benchmarking in wireless communication. These models use distributions that are easy to sample and therefore convenient for rapid benchmarking. They target generic scenarios relying on certain abstractions and simplifications. For instance, TDL channel models utilize pre-defined delay profiles with Rayleigh or Rician stochastic channel gains. Rayleigh fading model or Jakes model assume large number of arriving paths  uniformly distributed in the space. 

There are a few reasons for thinking differently about channel modeling. First, these models do not accurately capture the field data distribution. This mismatch would potentially incur performance loss if the communication system design is based on these models. This effect would become more pronounced as we try to leverage the power of machine learning based domain adaptation. More accurate models are crucial for demonstrating the gain expected from machine learning solutions. Second, it is cumbersome to build a channel model from field data measured in an environment. The current methods operate based on selecting a channel model and  fitting the parameters to the measured data. It is desirable to unify the model selection and parameter fitting steps and learn both in a systematic way. 

Recent advances in field of generative modeling made it possible to model more complicated data distributions and yet being able to flexibly sample from them. In this work, we propose a method that addresses the above issues based on generative modeling. The challenge is to design a generative model that captures well the main features of the communication channel such as delay profile and spatial correlation. 

We propose MIMO-GAN, a \ac{GAN} based model to directly learn the distribution of \ac{CIR}, denoted by $\vh$, only based on the transmit and receive signal pairs $(\vx,\vy)$. The generator models the distribution of channel taps. One challenge is that the ground truth \ac{CIR} is not given and therefore cannot be used to train the discriminator. Therefore, we can only train the discriminator on features extracted from $(\vx,\vy)$. For a generated \ac{CIR} from the model and knowing transmit signal $\vx$, we assume that the channel is linear and generate the receive signal using simple convolution. To learn specific features of \ac{CIR}, we first relate them to some receive signal features and train the discriminator on them. For example, the receive spatial correlation of the MIMO channel is learned by using $\vy^H\vy$ as input to the discriminator. We will comment more on the details of this design.

The resulting generative model can learn more complex distribution, and it is easier to sample from. Using neural networks to model the channel brings additional potential benefits. First, the neural implementation of the model and its parallel computation, and ease of sampling can bring simulation speed-ups.
With growing dimensionality of the channel, as in massive MIMO systems, or with increasing complexity of the modeling as in mmWave and Terraherz applications, the simulation phase becomes more time consuming. 
Although the conventional simulators can be accelerated with parallelized implementatinos, neural models are inherently  parallelizable and can reduce considerably the benchmarking rounds.
Second, these models are differentiable and can be used in  end-to-end designs of communication systems where it is required to back-propagate through the channel. 
Third, although the linearity assumption is used in our current design, we can use a non-linear mapping from $(\vx,\vh)$ to $\vy$ and learn non-linear effects caused by hardware impairments. Finally, we can mention that generative models can be used for better estimation and prediction of channel parameters similar to \cite{bora_compressed_2017} and follow up works.

The main contributions of the paper are as follows. We propose a \ac{GAN}-based model for statistical modeling of MIMO channels, that implicitly learns the distribution of band-limited time-domain impulse responses using input-output measurements. We show how to design the discriminator to learn spatial correlation for MIMO channels. By evaluating this approach on a dataset generated from 3GPP standard model, we demonstrate the consistency of MIMO-GAN with the underlying channel across a range of metrics e.g., delay profiles, spatial correlations. 

This paper is organized as follows. We review the related works in Section \ref{sec:related}. In Section \ref{sec:approach}, we present the main approach and architecture details. The numerical results are presented in Section \ref{sec:evaluation}.

\section{Related Work}
\label{sec:related}

A long line of literature tackles modelling the relation between wireless input-output signals and understanding influence of various factors (e.g., carrier frequency, relative movement) on communication; see the classic work of \cite{jakes_microwave_1994} for an example. 
Such works predominantly require significant domain expertise to construct mathematical channel models and is also further accompanied by extensive field measurements to identify parameters for these models.

Recently, there has been a focus on data-driven based approaches towards modelling channels, wherein the wireless channel model is represented by a deep neural network (DNN).
DNN-based channel models are appealing to rapidly build channel models that capture the data distribution better. They further enable differentiable computations and end-to-end learning of transceiver algorithms.
Existing approaches show promising results achieved by DNN-based channel models, such as  estimating the gradient of the channel using policy gradient \cite{aoudia_end--end_2018}, learning conditional output distributions of the channel over QAM symbols \cite{o2019approximating, ye2018channel, yang2019generative}, modelling AWGN, Rayleigh fading channels \cite{o2019approximating} and frequency-selective channels \cite{ye2020deep} and in some cases, also extending to collected measurement data \cite{dorner2020wgan}.
Common to these approaches is leveraging advances in unsupervised learning, such as by using generative adversarial networks (GANs) \cite{goodfellow2014explaining} or variational autoencoders (VAEs) \cite{kingma2013auto} to represent the channel model.

Although existing DNN-based channel model approaches have shown to be successful in certain scenarios, they are limited by some shortcomings which we address in this paper. First, to the best of our knowledge, the existing methods are limited mostly to SISO scenarios.
Second, it is unclear whether the approaches also extend to simulating complex channels (e.g., frequency-selective MIMO channels) as experimentally the approaches are largely evaluated in relatively simple scenarios (e.g., AWGN). We tackle this by proposing an approach that we show effective on 3GPP 4$\times$4 frequency-selective TDL MIMO channels. 

Next, unlike other statistical channel models, existing approaches model the conditional distribution specific to a particular setup (for example limited to QAM symbol communication), which restricts their usage as a general purpose simulation tool (e.g., for different constellations, modulation and MIMO dimensions). 
To address generalizing to novel inference scenarios, we propose learning the MIMO channel \ac{CIR} instead of general input-output mapping.

\section{Approach}
\label{sec:approach}

In this section, we begin by introducing the system model in \autoref{sec:approach_sysmodel} and then describe in detail our proposed approach MIMO-GAN in \autoref{sec:approach_mimogan}.

\subsection{System Model}
\label{sec:approach_sysmodel}

Consider a static MIMO channel with its discrete impulse response $\rmH(\tau)\in\C^{N_R\times N_T}$ at a fixed sampling frequency given by:
\begin{align}\rmH(\tau)&=
\begin{pmatrix}
h_{1,1}(\tau) & h_{1,2}(\tau) & \cdots & h_{1,N_T}(\tau) \\
h_{2,1}(\tau) & h_{2,2}(\tau) & \cdots & h_{2,N_T}(\tau) \\
\vdots  & \vdots  & \ddots & \vdots  \\
h_{N_R,1}(\tau) & h_{N_R,2}(\tau) & \cdots & h_{N_R,N_T}(\tau) \label{eq:channel_tensor}
\end{pmatrix}
\end{align}
where $\tau=0,\dots,L-1$, and $L$ is the number of taps. The transmit signal at sample $n$ with the sample sampling frequency is given by $\rvx[n]$. The receive signal is given by the discrete convolution:
\begin{align}
\rvy &= \rmH * \rvx \label{eq:channel_io_short}. 
\end{align}
We choose $T$ samples from the input and output of the channel. $T$ is chosen large enough to cover the communication duration. Note that the sampling frequency choice determines the maximum bandwidth we can simulate using the proposed generative model. We denote the transmit signal from $j$'th antenna as $\vx_j \in \sC^{T}$ and the receive signal from $i$'th antenna as $\vy_i \in \sC^{T}$.

\subsection{Proposed Model: MIMO-GAN}
\label{sec:approach_mimogan}

Having presented the system model in the previous section, we now elaborate on our approach by first presenting the key idea and then the details of our GAN-based approach.

\begin{figure*}
    \begin{center}
       \includegraphics[width=0.85\linewidth]{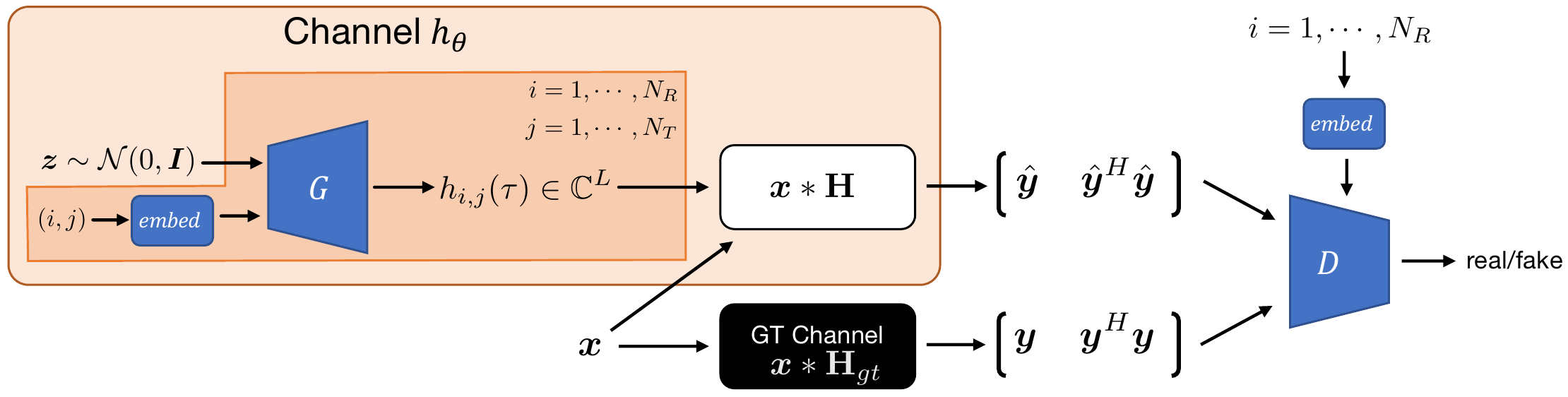}
    \end{center}
   \caption{\textbf{MIMO-GAN.} Approach overview.}
	\label{fig:mimogan_sys}
	\vspace{-1.0em}
\end{figure*}

\myparagraph{Motivation and Key Idea}
A recent line of work \cite{dorner2020wgan} has demonstrated reasonable success by \textit{explicitly} learning the input-output relation using deep neural networks (DNN).
Specifically, learning the relation is achieved by modeling $f$ as a DNN $f_\theta(\rvx, \vz)$ where the parameters $\theta$ are optimized to capture the channel effects and $\vz$ is used to introduce stochasticity.
While they have demonstrated promising results, we find explicitly learning the input-output relation leads to three shortcomings:
(a) it is unclear how $f_\theta$ (or generally the channel model) generalizes to certain configuration changes e.g., increasing the CP length and thereby the input length; such a configuration change requires $f_\theta$ to be retrained;
(b) the channel effects by default are not interpretable; and
(c) at inference time we want the model to generalize to any arbitrary input $\rvx \in \gX$. 
However, in practise it is challenging with only a limited set of training data to capture the diversity over the entire input space $\rvx \in \gX$ (lengthy waveforms in our case). 
This results in encountering out-of-distribution (OOD) data at inference time with a high probability, which has been shown by many works \cite{goodfellow2014explaining,hein2019relu} to result in erratic outputs.

The key idea in our work is to tackle these shortcomings by \textit{implicitly} learning the input-output relation.
Instead of modelling $f$ (the explicit input-output relation), we implicitly model the distribution of stochastic channel effects $p(\rmH)$ using a generative model $h_\theta$.
At inference time, we can sample instances of channel response filters $\rmH$ from the generative model and apply its effect on the input signal.
Since the channel filter is stochastically generated independent of the input, the learnt model has the benefit of being invariant to certain input-encoding factors and notably the duration of the input waveform.

We now provide the necessary background of Generative Adversarial Networks \cite{goodfellow2014generative}, a generative model which helps us capture stochastic channel effects. 
Following the background, we elaborate on the specifics of the MIMO-GAN approach.

\myparagraph{Background: Generative Adversarial Networks (GAN)}
Since our approach builds on top of building DNN-based generative models using the GAN framework \cite{goodfellow2014generative}, we provide the background necessary for the elaboration that follows.
The objective in GAN is to train a DNN model that captures the distribution $p(\rvx)$ of some observed data $\{\vx_i\}_{i=1}^N$, allowing one to sample new datapoints.
This is achieved by leveraging two ML models (typically DNNs): 
(i) a generator $G: \vz \mapsto \vx$ which transforms a random noise vector (drawn from $d$-dim isotropic Gaussian distribution) into a sample of the distribution; and
(ii) a discriminator $D: \vx \mapsto [0, 1]$ which predicts whether a given sample was generated by $G$ (i.e., considered `fake') or if it is a sample from the empirical distribution (i.e., considered `real').
The two networks $G$ and $D$ are trained simultaneously using a min-max objective:
\begin{align}
    \min_G \max_D \; & \E_{\vx \sim p_{\text{data}}} [\log D(\vx)] - \E_{\vz \sim p(\vz)}\left[ \log \left(D(G(\vz)) \right) \right] \label{eq:gan_vanilla}
\end{align}
While this formulation has proven successful, it typically suffers from training instabilities (which we also observed in our experiments) due to large variance in gradients during optimization.
To tackle this shortcoming, a recent line of work \cite{arjovsky2017wasserstein} propose enforcing a Lipschitz constraint, such as by using a gradient penalty term \cite{gulrajani2017improved}:
\begin{align}
    \min_G \max_D \; & \E_{\vx \sim p_{\text{data}}} [ D(\vx)] - \E_{\vz \sim p(\vz)}\left[ \left(D(G(\vz)) \right) \right] \nonumber \\
        & - \lambda \E_{\tilde{\vx}} \left[ \left( || \nabla_{\tilde{\vx}} D(\tilde{\vx}) ||_2 - 1 \right)^2 \right] \label{eq:wgangp}
\end{align}
where $\tilde{\vx}$ in the last term (gradient penalty) corresponds uniformly sampled interpolates between real $\vx \sim p_{\text{data}}$ and generated $\hat{\vx} \sim G(D(\vz))$ samples.
In this formulation, $D: \vx \mapsto \R$ is a value function and no longer maps inputs to probability scores.
Note that although $D$ is more precisely a `critic' in the WGAN formulation, we refer to it as a `discriminator' to remain consistent with GAN literature.

\myparagraph{MIMO-GAN: Overview}
The core idea of our approach is to implicitly model the distribution over band-limited impulse responses $p(h)$, which captures the stochastic channel effects.
It is implicit as we leverage only input-output measurements during training.
As detailed in \autoref{fig:mimogan_sys}, MIMO-GAN primarily involves two DNNs: a generator $G$ and discriminator $D$.
The generator $G$ is at the core of learnt channel $h_\theta$ (orange box of \autoref{fig:mimogan_sys}). 
To perform a simulation, we 
(a) sample an instance of MIMO channel $\rmH(\tau)$ from the generator $G$; and
(b) convolve the input signal $\vx$ with sampled instance.
We now take a closer look at each $G$ and $D$, and conclude the section with details on training.

\myparagraph{Generative Channel Model}
The backbone to the generative channel model $h_\theta$ is a DNN generator model $G$ that transforms a noise vector ($\vz \sim \gN(\bm{0}, \mI)$) into a sample from the distribution of interest ($p(h)$ in our case).
Specifically, the generator produces channel instances \textit{conditioned} on the spatial co-ordinates of the tx-rx antenna $(i, j)$.
The conditional co-ordinates, which are given as 1-hot encodings, are embedded into a $D$-dim space (set to 4 in our experiments) using a linear layer.
The conditional generation of channels based on embedded link-level information is particularly beneficial for three reasons.
First, since parameters of $G$ are shared for generation channels $\vh_{i, j}$ across all links $(i, j)$, the network is encouraged to learn and exploit correlations between the links.
Second, we have parameter efficiency since size of $G$'s parameters can remain constant in spite of large antenna arrays.
Third, we could leverage the same network for new configurations e.g., 8x8 (assuming critical spacing) can be used in e.g., 4x4, 8x2 setups.
To sample and construct a MIMO channel $\rmH$, we parallelize iterations over the spatial co-ordinates $(i, j)$ and build up the channel tensor (as shown in \autoref{eq:channel_tensor}).
With the sampled $\rmH$ and by treating the channel as an LTI system, we evaluate the output as \autoref{eq:channel_io_short}.

\myparagraph{Discriminator}
We leverage a discriminator DNN $D$ to encourage the generative channel model $h_\theta$, given some input $\rvx$, to produce realistic channel samples $\hat{\rvy} \in \sC^{N_R \times T}$ consistent with the GT channel $\rvy \in \sC^{N_R \times T}$ for a fixed input $\rvx \in \sC^{N_T \times T}$.
The goal of the discriminator is to weed out `fake' generated samples from `real' GT samples.
Similar to the generative channel model $h_\theta$, we also provide conditional information to the discriminator in the form of embeddings of receive co-ordinates $embed(i)$ via a linear layer.
As a result, the discriminator discriminates between real and fake samples of output waveforms $\vy, \hat{\vy}$ conditioned on the receive co-ordinate $i$.
The spatially-conditioned discriminator, similar to the conditional channel generation, benefits from evaluating in different MIMO configurations and the numbers of parameters remaining constant in different MIMO configurations.
To further complement the local (rx-specific) information observed in the samples, we additionally introduce global information (across all rx antennas) in the form of rx-side spatial correlations $\vy^H \vy \in \sC^{N_R \times N_R}$.

\myparagraph{Training}
We extend the original WGAN-GP objective (\autoref{eq:wgangp}) to implicitly learn the distribution over channels $\rmH \sim G(\vz)$.
However, since our goal is to learn this only using input-output measurements, we rewrite the objective in terms of the channel outputs:
\begin{align}
    \min_G \max_D \; & \E_{\vx \sim p_{\text{data}}} [ D(\vy, \vy^H\vy)] - \E_{\vz \sim p(\vz)}\left[ \left(D( \hat{\vy}, \hat{\vy}^H \hat{\vy}) \right) \right] \nonumber \\
        & - \lambda \E_{\tilde{\vy}} \left[ \left( || \nabla_{\tilde{\vy}} D(\tilde{\vy}, \tilde{\vy}^H \tilde{\vy}) ||_2 - 1 \right)^2 \right] \label{eq:mimogan_obj}
\end{align}
where $\hat{\vy} = \vx * \rmH$.
Note that unlike a typical GAN where the discriminator directly critiques the generator's outputs ($\rmH$ in this case), instead critiques the output of the channels $\hat{\vy}$.

\myparagraph{Implementation Details}
We model $G$ and $D$ as multi-layer perceptrons (MLPs) with ReLU activations, both two hidden layers of 100 units.
We train both networks simultaneously using Adam optimizer with a learning rate of 2$\times$10$^{-4}$ for 500 epochs.
For every training iteration of $G$, we make 25 iterations of the discriminator (critic) network $D$.
We use a gradient penalty weight $\lambda$ set to 10.

\section{Evaluation}
\label{sec:evaluation}

\subsection{Dataset}
\label{sec:eval_dataset}

To train and evaluate our generative channel model MIMO-GAN, we collect synthetic input-output measurements from a reference GT channel.
We now elaborate on the GT channel and details on measurements.

\myparagraph{Ground-truth Reference Channel}
We use as reference the 3GPP TDL channels \cite{3gpp_38901_2020} with a long delay spread (300 ns) and specifically TDL-A and TDL-B.
The TDL channels captures properties of noiseless channels with multipath fading demonstrating stochastic gains across a set of 23 paths.
For the MIMO setting, we consider co-polar uniform linear array antennas with $N_T$ and $N_R$ elements and `Medium-A' correlation.
In our experiments, we primarily consider $N_T$=4 and $N_R$=4.
We assume that the channel is stationary (by configuring the doppler shift to 0 Hz in simulations).
In particular, we use the Matlab implementation of the TDL channels to simulate input-output measurements.

\myparagraph{Channel Input-Output Measurements}
Using the reference GT TDL-A and TDL-B channels, we collect 60K input-output measurements at a bandwidth of 30.72 MHz.
The input signal per transmit antenna is fixed to a digital impulse signal with unit power.
We consider two approaches to transmit the input:
(i) by transmitting the digital impulse \textit{simultaneously} across all transmit antennas; and
(ii) by transmitting the impulse \textit{sequentially} over the transmit antennas.
By default, we use the latter approach of sequential transmissions, which as we show in the next section, allows us to faithfully capture transmit-side spatial correlations.
We sample the channel for a duration of 4.17 $\mu$s (thereby collecting 128 samples).
We find the sampling duration is sufficient to capture the last multipath component in our evaluation setting (2.89 $\mu$s for TDL-A and 1.43 $\mu$s for TDL-B).
As a result of this effort, we collect 60K input-output measurements, out of which we use 60\% for training and reserve 20\% each for validation and testing.

\subsection{Results}
\label{sec:eval_results}

In this section we walk through the evaluation of our MIMO-GAN approach trained and evaluated on 3GPP TDL MIMO channels.
Specifically, we evaluate the MIMO-GAN for the TDL-A and TDL-B channel models and compare the statistics with the underlying ground-truth channel.

\begin{table*}[t]
\centering
\vspace{3mm}
\caption{Power and delay statistics of MIMO-GAN and ground-truth (GT) channels.}
\label{tab:res_pdp}
\begin{tabular}{@{}lllll@{}}
\toprule
                       &          & Total Power (dB) & Average Delay ($\mu$s) & RMS Delay Spread ($\mu$s) \\ \midrule
\multirow{3}{*}{TDL-A} & MIMO-GAN & 4.648            & 0.2643             & 0.2862           \\
                       & GT       & 4.628            & 0.2641             & 0.2897           \\
                       & MAE      & -18.73           & 2.36$\times$10$^{-4}$           & 3.57$\times$10$^{-3}$         \\ \midrule
\multirow{3}{*}{TDL-B} & MIMO-GAN & 4.735            & 0.2276             & 0.2954           \\
                       & GT       & 4.688            & 0.2285             & 0.2987           \\
                       & MAE      & -14.95           & 9.05$\times$10$^{-4}$           & 3.37$\times$10$^{-3}$         \\ \bottomrule
\end{tabular}
\end{table*}

\begin{table*}[t]
\centering
\caption{MAEs of Power and Delay statistics comparing \textit{unconditioned} (`\xmark') and \textit{conditioned} (`\cmark') generation/discrimination of channels. 
Last row corresponds to MIMO-GAN. 
In each column, we represent the best performance in \textbf{bold}.
}
\label{tab:res_pdp_cond}
\begin{tabular}{ccllllllll}
\toprule
        &         &  & \multicolumn{3}{c}{TDL-A}             &  & \multicolumn{3}{c}{TDL-B}             \\ \cmidrule{4-6} \cmidrule{8-10}
Cond. G & Cond. D &  & Total power (dB) & Avg. Delay (ns) & RMS Spread (ns) &  & Total power (dB) & Avg. Delay (ns) & RMS Spread (ns) \\ \cmidrule{1-2} \cmidrule{4-6} \cmidrule{8-10}
\xmark       & \xmark       &  & -11.47      & 49.00     & 82.00      &  & \;-0.13            & 105.20          & 152.8          \\
\xmark       & \cmark       &  & -11.67      & 38.10     & 68.00      &  & \;-6.03            & \;82.94          & 133.4          \\
\cmark       & \xmark       &  & -12.25      & \;3.76      & \;7.93       &  & -11.74           & \;18.72          & \;23.01          \\
\cmark       & \cmark       &  & \textbf{-18.73}      & \;\textbf{0.24}      & \;\textbf{3.57}       &  & \textbf{-14.95}      & \;\;\textbf{0.90}   & \;\;\textbf{3.37}  \\ 
\bottomrule
\end{tabular}
\end{table*}

\begin{figure*}[t]
     \centering
     \begin{subfigure}[b]{0.48\linewidth}
         \centering
         \includegraphics[width=\textwidth]{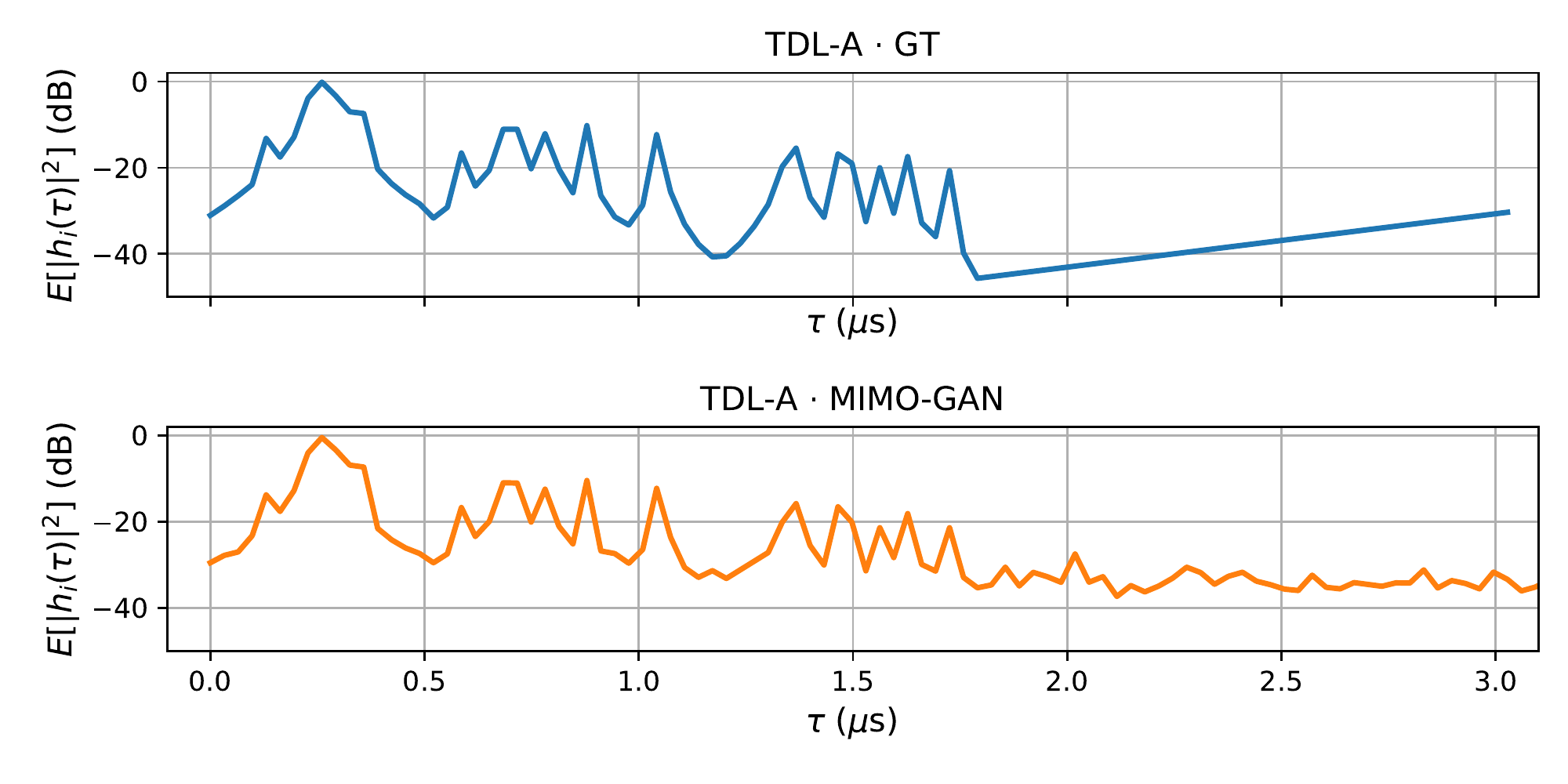}
         \label{fig:res_tdl_a_impulse_response_exp}
     \end{subfigure}
     \begin{subfigure}[b]{0.48\linewidth}
         \centering
         \includegraphics[width=\textwidth]{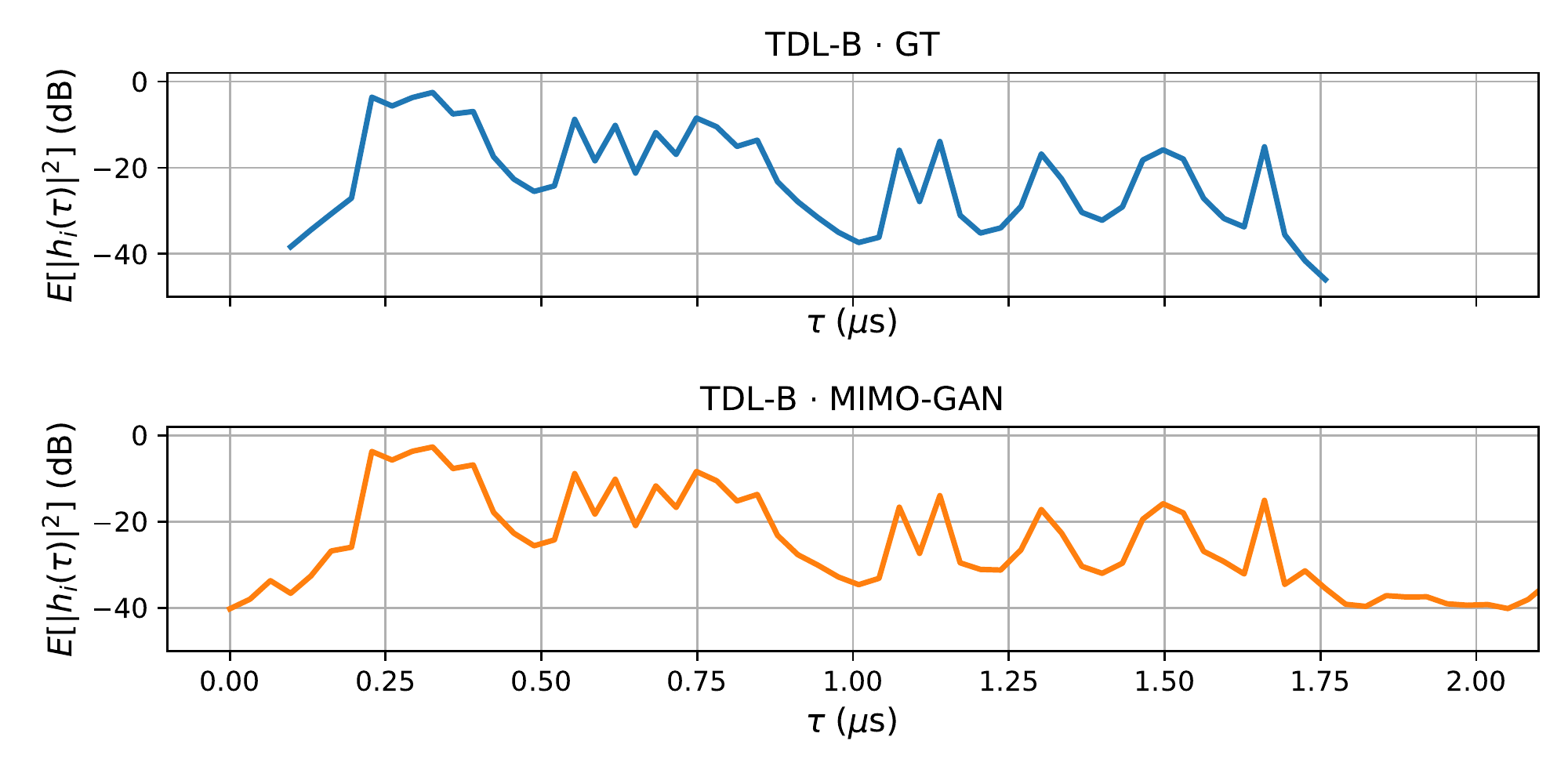}
         \label{fig:res_tdl_b_impulse_response_exp}
     \end{subfigure}
     \vspace{-2.0em}
    \caption{Power Density Profile (TDL-A and TDL-B)}
    \label{fig:res_tdl_pdp}
\end{figure*}

\begin{figure}[t]
    \begin{center}
       \includegraphics[width=0.9\linewidth]{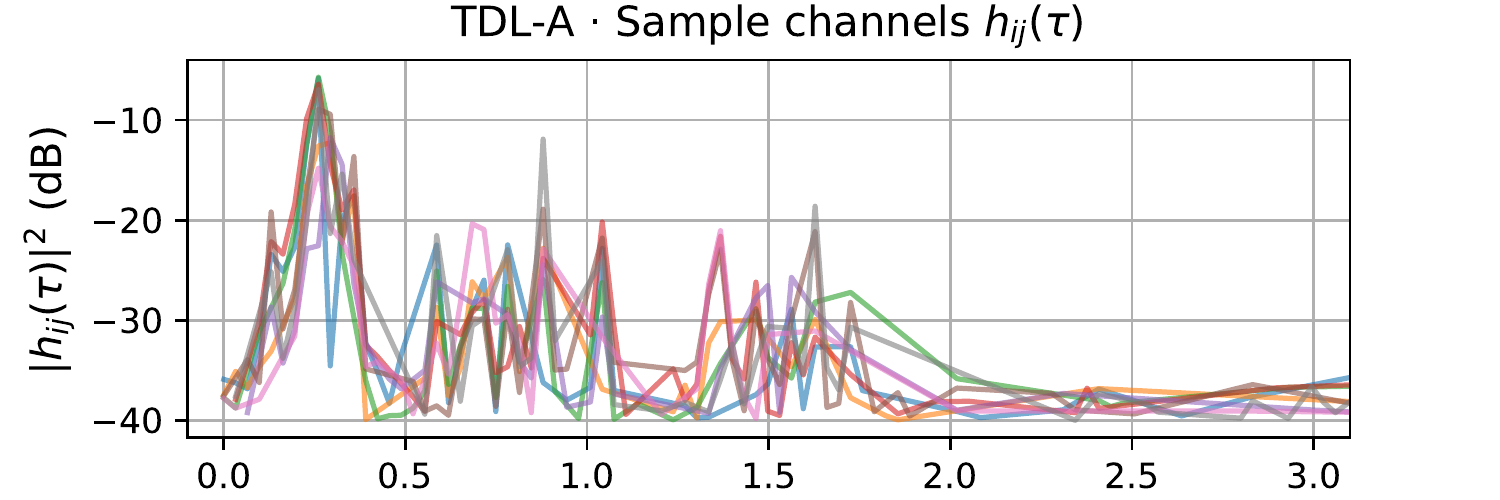}
    \end{center}
   \caption{Samples of channels $h_{ij}$ drawn from MIMO-GAN.}
	\label{fig:res_tdl_a_samples}
\end{figure}

\begin{figure}[t]
     \centering
     \begin{subfigure}[b]{0.9\linewidth}
         \centering
         \includegraphics[width=\textwidth]{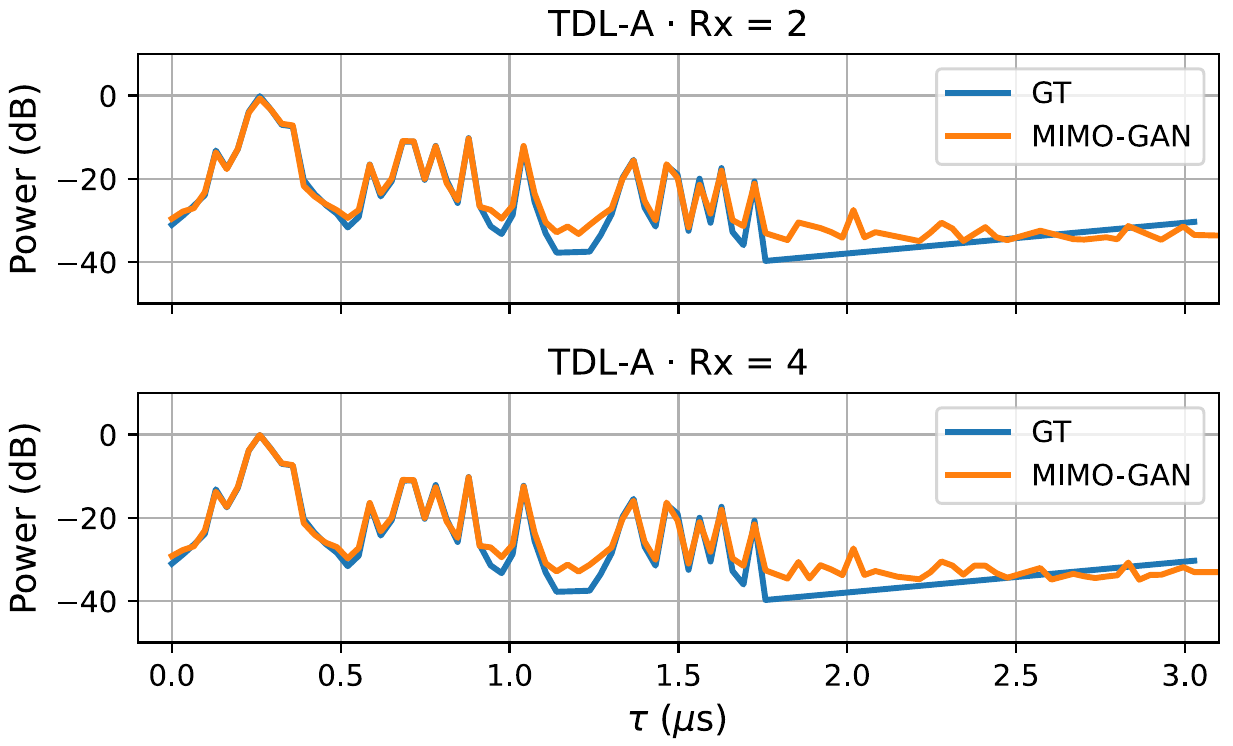}
         \label{fig:res_tdl_a_impulse_response}
     \end{subfigure}
     \begin{subfigure}[b]{0.9\linewidth}
         \centering
         \includegraphics[width=\textwidth]{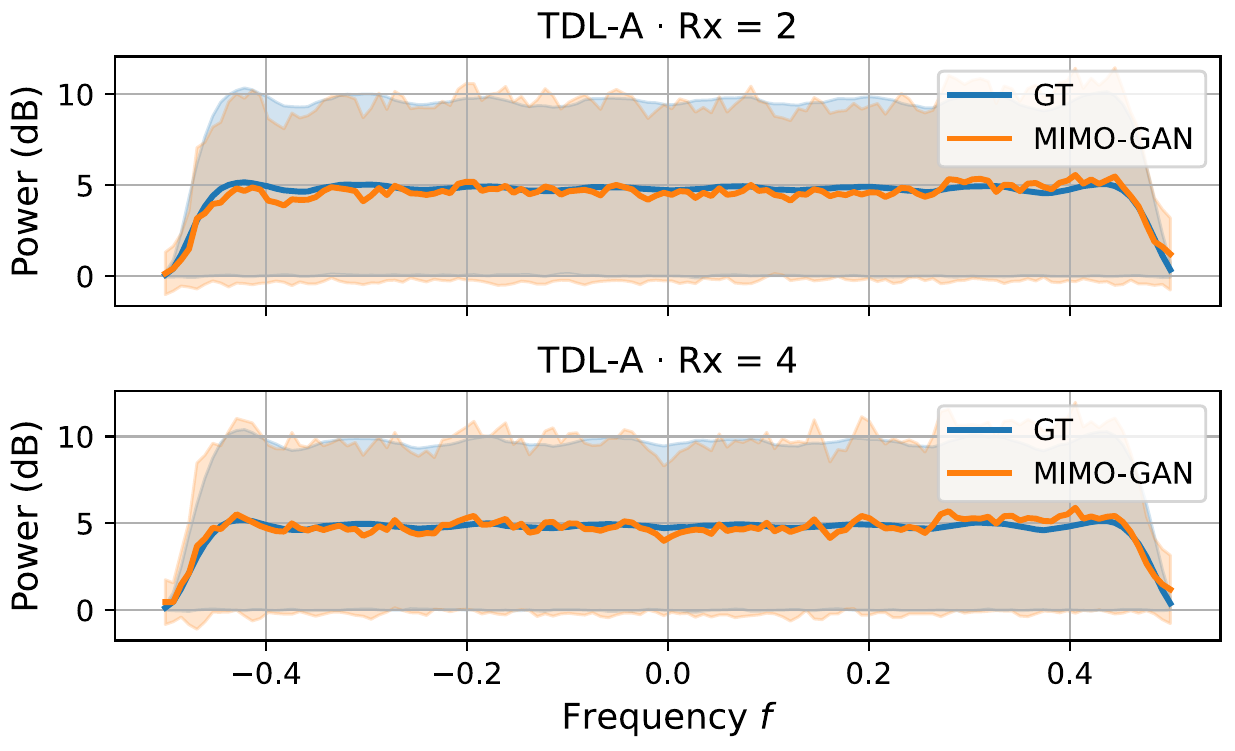}
         \label{fig:res_tdl_a_spectral_response}
     \end{subfigure}
        \caption{Power profile and Spectral density (TDL-A) in 4$\times$4 MIMO configuration (shown here for second and fourth antenna elements).}
        \label{fig:res_tdl_a_pdp_per_link}
\end{figure}

\myparagraph{Power Delay Profile}
We capture and compare the statistical parameters over a set of 12K impulse responses generated using MIMO-GAN and compare it with the parameters of the ground truth channel.
Specifically, we compare the statistical parameters defined in the ITU-R P.1407-8 \cite{itu_1407_2021} standards by using a power cutoff value of 20 dB below peak value.

In \autoref{tab:res_pdp}, we present the power delay statistics of our approach and ground-truth channels.
Across all metrics, we find that our approach can closely capture the GT statistics.
For instance, on TDL-A, we observe MIMO-GAN is within -18.69 dB mean absolute error (MAE) of power compared to the GT channel and within 3.57 ns of average delay.

Furthermore, in \autoref{fig:res_tdl_pdp}, we display the power delay profile of the impulse responses of both the GT channel and MIMO-GAN.
Remarkably, MIMO-GAN implicitly captures the impulse response of the learnt channel with reasonable accuracy; we observe mean absolute errors of -21.12 dB (TDL-A) and -20.69 dB (TDL-B).
We find these results also extend when observing across all receive antenna elements, as shown in \autoref{fig:res_tdl_a_pdp_per_link}.
As a result of implicitly modelling channe distribution using MIMO-GAN, we can also leverage the generator to sample channel instances, as shown in  \autoref{fig:res_tdl_a_samples}.

\myparagraph{Conditioned vs. Unconditioned $G$ and $D$}
In the previous section, we highlighted the consistency of MIMO-GAN with the underlying TDL-A and TDL-B channels when observing the power and delay statistics.
Now, we evaluate a key aspect of MIMO-GAN: learning the \textit{conditional} distribution of the channel, where the channel generation is conditioned on the spatial co-ordinates of the antenna element.
MIMO-GAN achieves this re-using the same network $G$ and $D$, but by additionally conditioning the network on spatial co-ordinates of tx-rx antennas $(i, j)$.
Now, we contrast the results by further examining an unconditional architecture, where:
(a) Generator $G$ generates the entire channel tensor $\rmH \in \C^{N_R\times N_T \times L}$ (as opposed to previously generating $\vh_{ij} \in \C^L$); and
(b) Discriminator $D$ discriminating the multiple-output signal $\rvy \in \sC^{N_R \times T}$ (as opposed to previously discriminating $\vy_i \in \C^T$.
Such an unconditional architecture is reminiscent of existing generative channel models approaches which have been primarily studied in a SISO setup.
Note that the unconditioned architectures introduces a significant increase in the number of trainable parameters (172K $\rightarrow$ 865K).

In \autoref{tab:res_pdp_cond}, we compare the results of MIMO-GAN (last row of table) with conditioned $G$ and $D$ with unconditioned counter-parts (first three rows of table).
Remarkably, we find that the conditional architecture results in performance improvements across all metrics in both TDL-A and TDL-B.
For instance, in TDL-A, we find an improvement of -12.37 dB (-11.47 dB vs. -18.73 dB) in total power and a 203$\times$ (49 ns vs. 0.24 ns) improvement in capturing average delay statistics.
We additionally investigate the influence of individually conditioning $G$ and $D$, as shown by the second and third rows of \autoref{tab:res_pdp_cond}.
Here, we find conditional \textit{generation} have a more significant impact on improving performance than conditional discrimination.
For instance, in TDL-A, conditional generation reduces MAE of average delays by 45.24 ns (92.3\% decrease) whereas conditional discrimination reduces MAE only by 10.9 ns (22.2\$ decrease).
We attribute the performance improvements to the networks being able to more efficiently exploit correlations of channels across various links.

\begin{figure}[t]
     \centering
     \vspace{3mm}
     \begin{subfigure}[b]{1.0\linewidth}
         \centering
         \includegraphics[width=\textwidth]{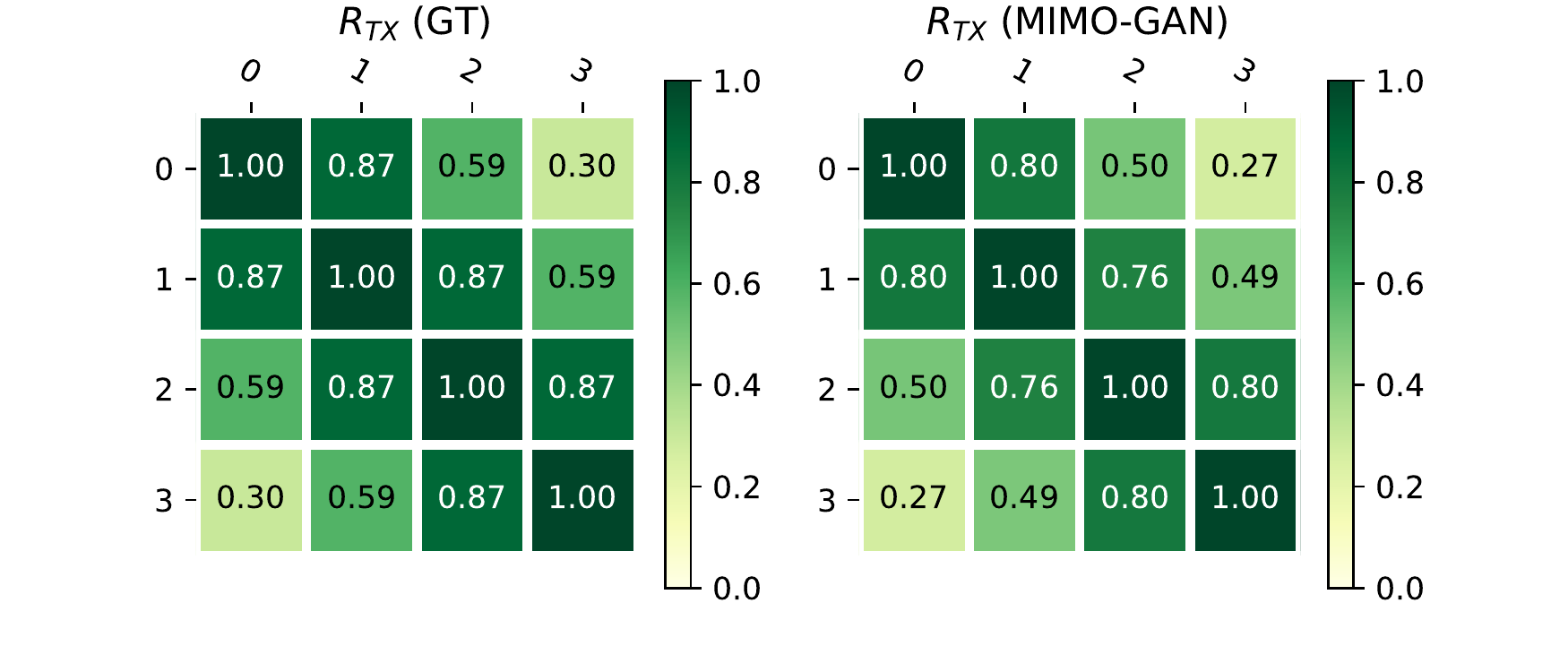}
         \caption{Transmit-side correlations $R_{TX} = \E[\rmH^H \rmH]$}
         \label{fig:res_tdl_a_spat_corr_tx}
     \end{subfigure}
     \begin{subfigure}[b]{1.0\linewidth}
         \centering
         \includegraphics[width=\textwidth]{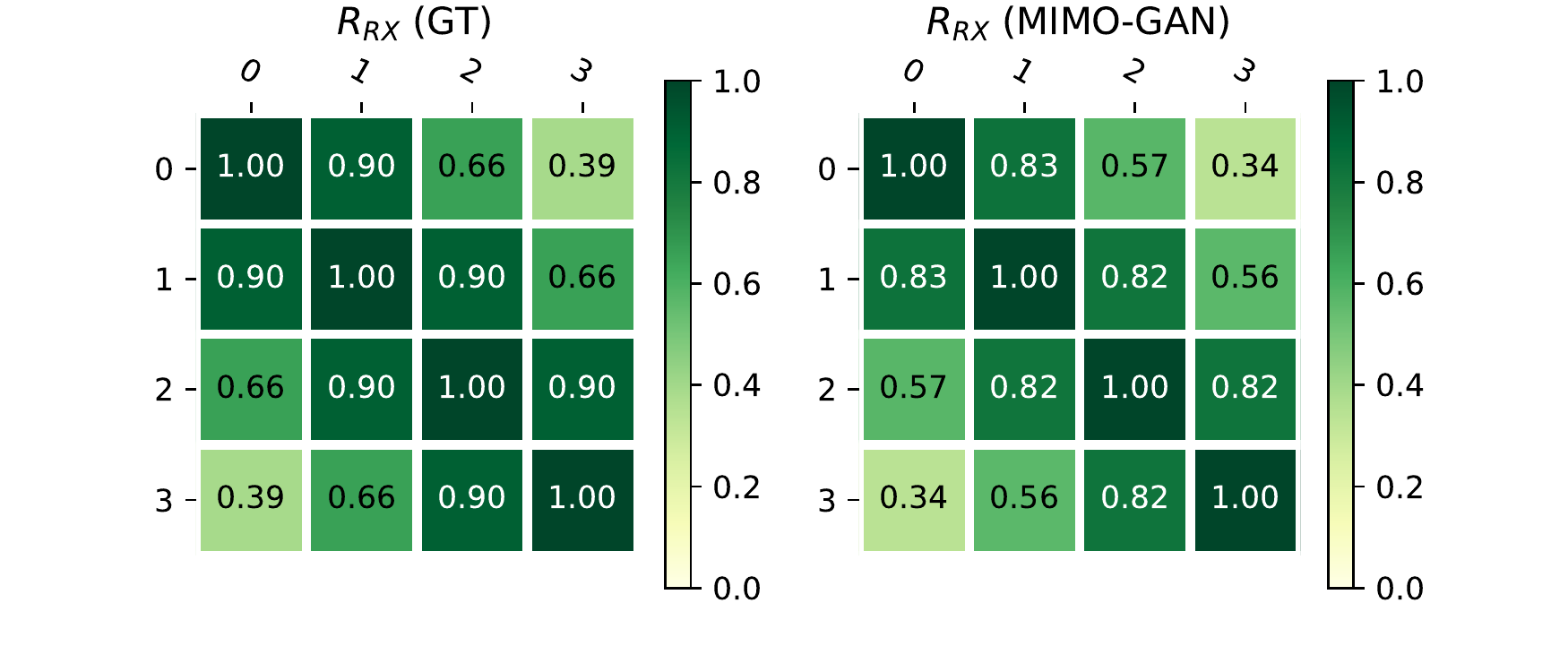}
         \caption{Receive-side correlations $R_{RX} = \E[\rmH \rmH^H]$}
         \label{fig:res_tdl_a_spat_corr_rx}
     \end{subfigure}
        \caption{Antenna Spatial Correlations in our 4$\times$4 MIMO setup. }
        \label{fig:res_tdl_a_spat_corr_txrx}
\end{figure}

\begin{table}[t]
\centering
\caption{Mean absolute errors between the GT and predicted antenna spatial correlations. 
`GM' indicates discriminating additionally on the receive-side gram matrices $\rvy^H \rvy$ (\autoref{eq:mimogan_obj}).
`SQ' indicates transmitting the digital impulse sequentially along the transmit antennas.
In each column, we represent the best performance in \textbf{bold} and second-best in \textit{italics}.
}
\label{tab:spat_corr}
\begin{tabular}{@{}ccllllll@{}}
\toprule
   &    & & \multicolumn{2}{c}{TDL-A} & & \multicolumn{2}{c}{TDL-B} \\ \cmidrule{4-5} \cmidrule{7-8}
GM & SQ & & $R_{TX}$     & $R_{RX}$     & & $R_{TX}$     & $R_{RX}$     \\ \cmidrule{1-2} \cmidrule{4-5} \cmidrule{7-8}
\xmark  & \xmark  & & 0.144       & 0.171       & & 0.139       & 0.143       \\
\cmark  & \xmark  & & 0.204       & \textit{0.061}       & & 0.154       & \textit{0.056}       \\
\xmark  & \cmark  & & \textit{0.071}       & 0.177       & & \textit{0.067}       & 0.164       \\
\cmark  & \cmark  & & \textbf{0.059}       & \textbf{0.057}       & & \textbf{0.058}       & \textbf{0.063}       \\ \bottomrule
\end{tabular}
\vspace{-1.0em}
\end{table}

\myparagraph{Spatial Correlations}
For MIMO evaluation, we also consider the transmit- and receive-sided spatial correlations of responses between antenna elements.
We display the correlations across top 10 taps for TDL-A channels in \autoref{fig:res_tdl_a_spat_corr_txrx}.
Overall, we observe promising results with mean absolute errors between the GT and predicted spatial correlations: 0.057 (TDL-A) and 0.063 (TDL-B).

We find two steps especially crucial to successfully capture the MIMO spatial correlations, as shown in \autoref{tab:spat_corr}.
First, to effectively capture the \textit{receive}-side correlations in the generated MIMO channels, we find it crucial to additionally condition the discriminator on the gram matrices ($\rvy^H \rvy$) of output measurements.
We find that including this conditioning (second row in \autoref{tab:spat_corr}) reduces the errors by 60-64\%.
Second, to capture the \textit{transmit}-side correlations, we observe it is essential to sequentially transmit the inputs over each transmit antenna.
Without this step, the receiver observes only the superposition of the impulses and is thereby ill-posed to learn channel conditions at the transmit side.
As a result of sequential input transmissions (third row in \autoref{tab:spat_corr}), we can similarly reduce the transmit correlation errors by 51\%.
Overall, with the above steps, we find that MIMO-GAN additionally captures transmit- and receive-side antenna spatial correlations with mean absolute errors of 0.057-0.063 (final row in \autoref{tab:spat_corr}).

\begin{table}[]
\centering
\caption{Simulation times. Speed-ups over GT TDL-A implementation.}
\label{tab:simulation_times}
\begin{tabular}{@{}lcclr@{}}
\toprule
\multicolumn{5}{c}{1$\times$1}                                    \\ \midrule
          & Batched & GPU & simulation time (ms)                    & speed up \\ \cmidrule{2-5}
GT    &  \xmark       & \xmark    & 9.67 $\pm$ 55.69   & 1.00     \\ \cmidrule{2-5}
\multirow{4}{*}{MIMO-GAN}  &  \xmark       & \xmark     & 0.53 $\pm$ 0.12    & 18.25    \\
          &  \cmark       & \xmark     & 0.11 $\pm$ 0.01    & 87.91    \\
          &  \xmark       & \cmark     & 0.99 $\pm$ 0.08    & 9.77     \\
          &  \cmark       & \cmark     & 0.02 $\pm$ 0        & 483.50   \\  \midrule
\multicolumn{5}{c}{2$\times$2}                                    \\  \midrule
          & Batched        & GPU    & simulation time (ms)  & speed up \\  \cmidrule{2-5}
GT    & \xmark        & \xmark     & 10.37 $\pm$ 54.39   & 1.00     \\ \cmidrule{2-5}
\multirow{4}{*}{MIMO-GAN}  & \xmark        & \xmark     & \ \,1.40 $\pm$ 0.27     & 7.41     \\
          & \cmark        & \xmark     & \ \,0.21 $\pm$ 0.02    & 49.38    \\
          & \xmark        & \cmark     & \ \,2.23 $\pm$ 0.06    & 4.65     \\
          & \cmark        & \cmark     & \ \,0.14 $\pm$ 0.01     & 74.07    \\  \midrule
\multicolumn{5}{c}{4$\times$4}                                    \\  \midrule
          & Batched        & GPU    & simulation time (ms) & speed up \\  \cmidrule{2-5}
GT    & \xmark        & \xmark     & 12.49 $\pm$ 56.07  & 1.00     \\ \cmidrule{2-5}
\multirow{4}{*}{MIMO-GAN}  & \xmark        & \xmark     & \ \,3.91 $\pm$ 0.62    & 3.19     \\
          & \cmark        & \xmark     & \ \,0.43 $\pm$ 0.04    & 29.05    \\
          & \xmark        & \cmark     & \ \,6.10 $\pm$ 0.16      & 2.05     \\
          & \cmark        & \cmark     & \ \,0.23 $\pm$ 0.01    & 54.30    \\ \bottomrule
\end{tabular}
\vspace{-1.0em}
\end{table}

\myparagraph{Simulation Times}
A side-benefit provided by neural surrogates for channel model, such as by MIMO-GAN, is enabling rapid simulations.
Such rapid simulations makes MIMO-GAN especially appealing for end-to-end training approaches \cite{raj2018backpropagating}; such approaches typically require hundreds of thousands of forward (simulations) and backward passes through the channel model.
We present the simulation times of TDL-A using MIMO-GAN in \autoref{tab:simulation_times} and contrast it with a publicly available implementation (Matlab in this case).
The table reports the mean and standard deviations of 2048 simulations of a time-invariant TDL-A channel.
Overall (last rows in the table), we observe that MIMO-GAN simulations are significantly faster, with speed-ups of 54.3-483$\times$.
We attribute these speed-ups to a combination of:
(a) batching (independently leading to 29-88$\times$ speed-ups), as multiple simulations (which in our case is a simple forward pass) can efficiently parallelized; and
(b) GPU usage (independently leading to 2-10$\times$ speed-ups), as the effectiveness of parallelized operations can be further amplified using GPU cores.

\subsection{Discussion}
In this paper, we proposed MIMO-GAN as a data-driven neural surrogate for channel models.
They key idea of MIMO-GAN is to implicitly learn the channel distribution $p(\rmH)$ from input-output measurements.
We discussed in \autoref{sec:eval_results} that in addition to faithfully capturing the power delay profile of the underlying channel, MIMO-GAN also offers many other benefits such as capturing the MIMO spatial correlations and rapid simulation times.

We envision MIMO-GAN, or more generally neural channel models, are appealing for many applications in wireless communications.
We highlight two such applications.
First, neural channel models allow end-to-end learning \cite{o2017introduction,dorner2017deep} of the physical layer.
In this case, parameters and operational configurations on both the transmitter and receiver can be optimized in response to the channel conditions experienced by the transceiver.
Second, given that MIMO-GAN depends on input-output measurement data (which can be collected on-device with relative ease), the model can be rapidly built on-device and further fine-tuned based on new measurement data.
Having access to such a neural channel model on-device opens up many possibilities, such as performing inference \cite{donahue2017adversarial} to obtain a compact channel representation.

While we found promising results with MIMO-GAN, our findings also highlights many interesting open challenges.
One such challenge is additionally modelling time-variation.
Although MIMO-GAN can be fine-tuned periodically to obtain instantaneous channel representations, there is scope for a more principled approach to model time-variation.
Advances in neural stochastic processes \cite{deng2020modeling} might hold the key to model time-variations in statistical channels.
Another challenge is extending the approach to massive MIMO systems.
While MIMO-GAN provides the benefit of parameter efficiency in such systems, it is hindered by computation complexity due to conditional sequential generation of channels.
We believe significant speed-ups, as to what might be needed for massive MIMO, can be obtained by efficient parallelized implementation of the generative process.

\section{Conclusion}
\label{sec:conclusion}

In this paper, we proposed `MIMO-GAN' towards building frequency-selective MIMO channel models based on input-output measurements.
The key benefits to our approach, as a result of implicitly learning the channel distribution, are interpretability (since instantaneous channels are visible) and being able to generalize to configuration changes (e.g., sampling the channel for variable durations).
Furthermore, 
Based on evaluations on 3GPP TDL channels, we found MIMO-GAN consistently captures the power delay statistics and additionally spatial correlations between antennas.

\bibliographystyle{IEEEtran}
\bibliography{references.bib}
\end{document}